\documentstyle[twoside,fleqn,espcrc2]{article}

\newcommand{\AmS}{{\protect\the\textfont2
  A\kern-.1667em\lower.5ex\hbox{M}\kern-.125emS}}

\title{Topological properties of QCD with two dynamical fermions}

\author{B. All\'es\thanks{Speaker at the conference.}\address{Dipartimento 
                           di Fisica,
                           Universit\`a di Milano--Bicocca and INFN
                           Sezione di Milano, Milano, Italy},
        M. D'Elia$^{\rm b}$, A. Di Giacomo\address{Dipartimento
                  di Fisica, Universit\`a di Pisa, Via Buonarroti 2,
                  Ed. B, 56127 Pisa, Italy}}
\begin{document}

\begin{abstract}
We investigate the topological susceptibility
of the QCD vacuum with two flavours of
dynamical staggered fermions on the lattice
both at zero and finite temperature. At zero temperature we study the
dependence of the signal on the fermion mass and at finite temperature
we analyze the behaviour across the
phase transition.
\end{abstract}

\maketitle

\section{INTRODUCTION}

The topological susceptibility is an important parameter of the QCD
vacuum. In the continuum it is defined as
\begin{equation}
 \chi \equiv \int d^4x \;\; \partial_\mu 
   \langle 0 | {\rm T}\left\{K_\mu(x) Q(0)\right\}| 0 \rangle \; ,
\label{eq:chi}
\end{equation}
where $K_\mu(x)$ is the Chern current
\begin{eqnarray}
 K_\mu(x) &=& {g^2 \over 16 \pi^2} \epsilon_{\mu\nu\rho\sigma} \times \nonumber \\
  & &A_\nu^a \left( \partial_\rho A^a_\sigma - {1 \over 3} 
  g f^{abc} A_\rho^b A_\sigma^c \right) \; ,
\end{eqnarray}
and $Q(x)=\partial_\mu K_\mu(x)$ is
the density of topological charge,
\begin{equation}
 Q(x) = {g^2\over 64\pi^2} \; \epsilon^{\mu\nu\rho\sigma} \;
        F^a_{\mu\nu}(x) \; F^a_{\rho\sigma}(x)  \; .
\end{equation}
Eq.~(\ref{eq:chi}) uniquely defines the prescription for the singularity 
of the time ordered product when\break $x \rightarrow 0$~\cite{witten}.

We present results on the topological susceptibility in QCD with
$N_f=2$ degenerate flavours. 

\subsection{The Simulation}

\begin{figure}[htb]
\vspace{4.5cm}
\includegraphics{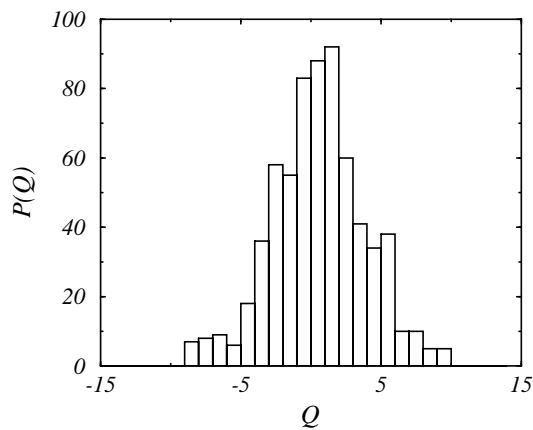}
\null\vskip 0.3cm
\caption{Distribution of topological charge after 30 cooling
steps at $\beta=5.5$.}
\label{fig:qdecorrel5.5}
\end{figure}

We have simulated the theory on a $32^3\times 8$ lattice with two flavours
of staggered fermions with bare mass $am=0.0125$. The R--type HMC
algorithm has been utilized for the updating~\cite{gottlieb}.
Each trajectory consisted in
60 steps of $\Delta\tau=0.005$ units of molecular dynamics time.
We have performed the simulation at the following values of the 
lattice bare coupling
$\beta\equiv 6/g^2=5.4$, 5.5, 5.6 and 5.7$\;.$
To be sure that the topology is well thermalized we have checked the 
topological charge of our sample of configurations by
cooling~\cite{boyd,lippert}.
In Fig.~\ref{fig:qdecorrel5.5} we show the 
distribution of topological charge from our configurations at 
$\beta=5.5\;$. The histogram displays a good sampling.
To achieve the same ergodicity at larger values of $\beta$ 
we had to increase the statistics. 

We have measured the Polyakov loop at zero and finite temperatures.
With our bare fermion mass and lattice size the deconfining 
transition seems to occur at $\beta_c = 5.54(2)$~\cite{bitar}. 

%\vskip 6mm

%\begin{figure}[htb]
%\vspace{4.5cm}
%\special{psfile=histo5.70_T.ps angle=-90 hscale=32 vscale=35 voffset=170
%hoffset=-10}
%\null\vskip 0.3cm
%\caption{Distribution of topological charge $Q_L$ after 30 cooling
%steps at $\beta=5.7$ and finite temperature.}
%\label{fig:qdecorrel5.7}
%\end{figure}

%
\begin{table*}[hbt]
% space before first and after last column: 1.5pc
% space between columns: 3.0pc (twice the above)
\setlength{\tabcolsep}{3.5pc}
% -----------------------------------------------------
% adapted from TeX book, p. 241
\newlength{\digitwidth} \settowidth{\digitwidth}{\rm 0}
\catcode`?=\active \def?{\kern\digitwidth}
% -----------------------------------------------------
\caption{Topological susceptibility, $T/T_c$ and $a^2\sigma$ vs. $\beta$.}
%and lattice spacing (the
%values of $a$ marked with an asterisk have been obtained from interpolation.)} 
\label{tab:tab}
\begin{tabular*}{\textwidth}{@{}l@{\extracolsep{\fill}}rrrr}
\hline
%                 & \multicolumn{2}{l}{Pilot plant} 
%                 & \multicolumn{2}{l}{Full scale plant} \\
%\cline{2-3} \cline{4-5}
%                 & \multicolumn{1}{r}{$a$/fm}
                 & \multicolumn{1}{r}{$\beta$}
                 & \multicolumn{1}{r}{$a^2\sigma$}
                 & \multicolumn{1}{r}{$T/T_c$}
                 & \multicolumn{1}{r}{$10^{-8}\;\chi$/MeV$^4$}  \\
\hline
%5.4 & 0.310(40) & 0.9677 & 1.61(43) \\
%5.5 & 0.306(30)(*) & 0.9804 & 1.13(30) \\
%5.6 & 0.300(20) & 1.0000 & 1.21(30) \\
%5.7 & 0.296(18)(*) & 1.0126 & 2.89(1.46) \\
%5.06 & 0.292(14)(*) & 1.0274 & 1.63(1.22) \\
%5.10 & 0.270(8) & 1.1110 & 0.43(1.19) \\
&5.4 & 0.138(8) &   0.707    &  5.16(1.66) \\
&5.5 & 0.082(9) &   0.917    &  6.21(1.74) \\
&5.6 & 0.053(3) &   1.143    &  2.27(0.72) \\
&5.7 & --       &   1.430    &  1.72(1.00) \\
\hline
%\multicolumn{5}{@{}p{120mm}}{Reprinted from: G.M. Ritcey,
%                             Tailings Management,
%                             Elsevier, Amsterdam, 1989, p. 635.}
\end{tabular*}
\end{table*}

\subsection{The Operators and their Renormalizations}

We have measured the topological charge density on the lattice
by using the operator $Q_L(x)$~\cite{haris}
\begin{equation}
 Q_L(x) = -{1 \over 2^9 \pi^2} \;\widetilde{\epsilon}_{\mu\nu\rho\sigma}
          {\rm Tr}\left( \Pi_{\mu\nu}(x) \Pi_{\rho\sigma}(x) \right)
          \; ,
\label{eq:ql}
\end{equation}
and applying two smearing steps on it.
The lattice topological susceptibility is defined by
\begin{equation}
 \chi_L \equiv {\langle Q_L^2 \rangle\over V} \; ,
\label{chilatt}
\end{equation}
where $Q_L$ is the total topological charge and $V$ the space--time
volume. To match this definition to the one given in Eq.(\ref{eq:chi})
we must subtract the singularity in the product of the two operators
in Eq.(\ref{chilatt}). We make the subtraction by using the
field--theoretical method~\cite{tanti} 
\begin{equation}
 \chi_L = Z^2 a^4 \chi + M \; ,
\end{equation}
and evaluating the multiplicative and additive renormalizations $Z$
and $M$ by use of the heating method~\cite{tanti2}. $M$ contains
mixings with gluonic and fermionic condensates. The lattice
topological charge $Q_L$ mixes with fermionic operators during
renormalization but it has been shown that the off--diagonal mixing is
negligible~\cite{panag}.

The heating method to evaluate $Z$ and $M$ yields a non--perturbative 
determination of these renormalization constants~\cite{tanti2,tsu3}.
To calculate $Z$ a local updating algorithm 
is applied on a configuration containing 
a charge +1 classical instanton. These updatings, being local, thermalize
the short distance fluctuations, responsible for the renormalization effects,
and due to the slowing down leave large structures, like instantons, unchanged.
The measurement of $Q_L$ on such updated configurations yields $Z \cdot Q$. As
$Q$ is known, one can extract $Z$. Notice that this procedure is equivalent
to imposing the continuum value for the 1--instanton charge (in the
$\overline{\rm MS}$ scheme it is +1) and extracting the finite multiplicative
renormalization $Z$ by evaluating a matrix element of $Q_L$.

The additive renormalization $M$ is obtained in a similar way. We
apply a few heating steps with a local updating algorithm on a zero--field
configuration. Then the topological susceptibility is calculated.
This provides the value of $M$, if no instantons have been created during
the few updating steps. This
method agrees with the treatment of the singularity of~Eq.(\ref{eq:chi}).

As explained in~\cite{tsu3}, cooling tests must be done to check that
the background topological charge has not been changed during the
local heating. 

\subsection{Determination of the scale}

The lattice spacing was extracted by measuring the string tension 
at $\beta=5.4$, 5.5 and 5.6 on a $16^4$ lattice and assuming
$\sqrt{\sigma}=420$ MeV. For $\beta=5.7$ it was determined 
by an extrapolation using the 2--loop beta function.
Wilson loops were evaluated by using smeared spatial links. 
The determinations of $a^2\sigma$ are shown in 
Table~\ref{tab:tab}. From the value at $\beta_c$ 
($a = 0.123(6)\;\; {\rm fm}$) we infer the critical 
temperature to be $T_c=200\pm 10\pm 15$ MeV where the first error is
our statistical one and the second comes from the indetermination in
the result for $\beta_c$.

\section{RESULTS}

\begin{figure}[htb]
\vspace{4.5cm}
\includegraphics{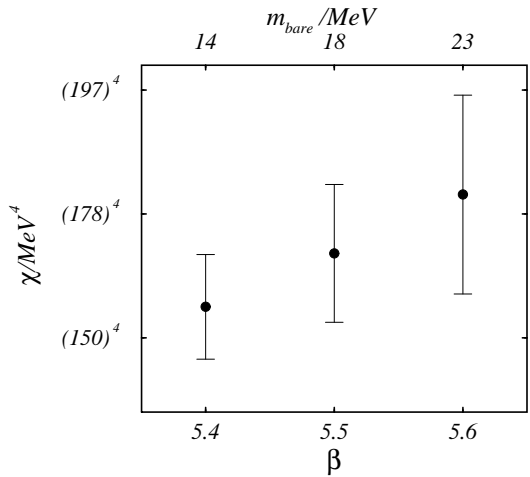}
\null\vskip 0.3cm
\caption{Topological susceptibility as a function
of the bare fermion mass.}
\label{fig:chiT0}
\end{figure}

The topological susceptibility at zero temperature is shown in
Fig.~\ref{fig:chiT0}. In the lower (upper) axis we show the values of
$\beta$ (bare fermion mass). It displays a trend
consistent with the theoretical expectation 
$\chi\propto m \langle\overline{\psi} \psi\rangle $.
%\begin{equation}
% \chi \propto m \langle\overline{\psi} \psi\rangle \; ,
%\end{equation}
%for small $m$.

\begin{figure}[htb]
\vspace{4.5cm}
\includegraphics{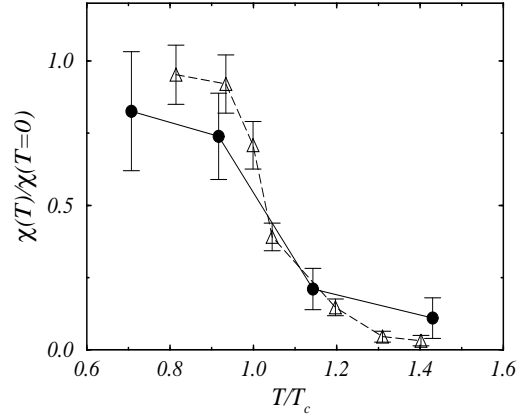}
\null\vskip 0.3cm
\caption{Behaviour of the topological susceptibility as a function
of the normalized temperature $T/T_c$. Black dots correspond to
$N_f=2$ and white triangles to $N_f=0$~\cite{tsu3}. Lines are drawn to
guide the eye.}
\label{fig:chiT}
\end{figure}

The values for the topological susceptibility at finite temperature 
are given in Table~\ref{tab:tab}.
In Fig.~\ref{fig:chiT} we show the normalized topological
susceptibility $\chi(T)/\chi(T=0)$ as a function of 
the temperature. The value at zero temperature for $\beta=5.7$ has
been obtained by extrapolating the results shown in
Fig.~\ref{fig:chiT0}. 
In Fig.~\ref{fig:chiT} the results for the quenched
case~\cite{tsu3} are shown for comparison.
The signal for $\chi$ drops when crossing the 
transition temperature. Within the large errors, 
the drop seems to be as sharp as it was for
the quenched case. Our results for $N_f=2$ are in qualitative 
agreement with the ones shown in~\cite{deforcrand}. For $N_f=4$ the
drop in the signal for $\chi$ turns out to be much
steeper~\cite{lat98}. 

\vskip 6mm

%\section{Acknowledgements}

%A. Di Giacomo acknowledges the financial contribution of the 
%European Commission under the TMR-Program ERBFMRX-CT97-0122.

\end{document}